\documentclass[12pt]{article}
\usepackage[cm]{fullpage}
\usepackage{amsmath}
\usepackage{amsfonts}
\usepackage{amssymb}
\usepackage{amsthm}
\usepackage{graphicx}
\usepackage{fancyhdr}
\usepackage{xcolor}

\definecolor{bb}{rgb}{0.3, 0.5, 1}
\definecolor{bg}{rgb}{0.1, 0.1, 0.5}
\usepackage[%
    pdftitle={The title of your letter},
    pdfauthor={Your name},
    pdfsubject={The subject of the letter},
    pdfkeywords={Any keywords},
     colorlinks=true,  
     linkcolor=bg, 
     citecolor=bg, 
     urlcolor=bg, 
     pdfnewwindow=true,
     pagebackref=true,
     filecolor=bb,
     pdffitwindow=false,     
     pdfstartview={FitH}    
]{hyperref}

\def\ba{\begin{eqnarray}}
\def\ea{\end{eqnarray}}
\def\be{\begin{equation}}
\def\ee{\end{equation}}

\def\nn{\nonumber}

\def\mn{_{\mu \nu}}

\def\({\left(}
\def\){\right)}
\def\ie{{\it i.e. }}
\def\nn{\nonumber}
\def\p{\partial}

\def\stu{St\"uckelberg }
\def\mpl{M_{\rm Pl}}

\begin{document}

\vskip 0.9cm

\centerline{\Large \bf Cosmology of the Galileon from Massive Gravity}
\vskip 0.7cm
\centerline{\large Claudia de Rham\footnote{Claudia.deRham@unige.ch} and Lavinia Heisenberg\footnote{Lavinia.Heisenberg@unige.ch}}
\vskip 0.3cm

\centerline{\em D\'epartement de Physique Th\'eorique and Center for Astroparticle Physics, }
\centerline{\em Universit\'e de Gen\`eve, 24 Quai E. Ansermet, CH-1211,  Gen\`eve, Switzerland}

\vskip 1.9cm

\begin{abstract}
We covariantize the decoupling limit of massive gravity proposed in \cite{deRham:2010kj} and study the cosmology of this theory as a proxy, which embodies key features of
the fully non-linear covariant theory. We first confirm that it exhibits a self-accelerating solution,  similar to what has been found in \cite{deRham:2010tw}, where the Hubble parameter corresponds to the graviton mass. For a certain range of parameters fluctuations relative to the self-accelerating background are stable and form an attractor solution. We also show that a degravitating solution can not be constructed in this covariantized proxy theory in a meaningful way. As for cosmic structure formation, we find that the helicity-0 mode of the graviton causes an enhancement relative to $\Lambda$CDM.
For consistency we also compare proxy theories obtained starting from different frames in the decoupling limit and discuss the possibility of obtaining a non-representative proxy theory by choosing the wrong starting frame.

\end{abstract}

\vspace{1cm}

\section{Introduction}

Observations of the CMB, supernovae, lensing and baryon acoustic oscillations have led to the cosmological standard model which requires an accelerated expansion of the late Universe, driven by dark energy but despite many years of research its origin has not yet been identified. There are two major explanations for the origin and properties of dark energy.

The first solution consists of introducing a cosmological constant $\Lambda$ with a constant energy density causing an effective repulsive force between cosmological objects at large distances. From the particle physics perspective the cosmological constant could correspond to the vacuum energy density.
The theoretical expectations for the vacuum energy density caused by fluctuating quantum fields, however, exceeds the observational bounds on $\Lambda$ by up to 120 orders of magnitude. This discrepancy remains for more almost a century one of the most challenging puzzles in physics, \cite{Straumann:2002tv}.

Alternatively, the acceleration of the Universe can be explained by introducing new dynamical degrees of freedom, either by invoking new fluids $T_{\mu\nu}$ with negative pressure or by changing the geometrical part of Einstein's equations. In particular, weakening gravity on cosmological scales could not only tackle the cosmological constant problem, but would also come hand in hand with new degrees of freedom which might be responsible for a late-time speed-up of the Hubble expansion. Such scenarios could arise in massive gravity or in higher-dimensional frameworks.

In the higher dimensional picture, the Dvali-Gabadadze-Porrati (DGP) model is one of the important large scale modified theories of gravity \cite{DGP}. In this braneworld model our Universe is confined to a three-brane embedded in a five-dimensional bulk. On small scales, four-dimensional gravity is recovered due to an intrinsic Einstein Hilbert term sourced by the brane curvature, whereas on larger, cosmological scales gravity is systematically weaker as the graviton acquires a soft mass $m$ which limits its effective range. Being a fundamentally higher dimensional theory, the effective four-dimensional graviton on the brane carries five degrees of freedom, namely the usual helicity-2 modes, two helicity-1 modes and one helicity-0 mode. Whilst the helicity-1 mods typically decouple, the helicity-0 one can mediate an extra fifth force.
In the limit $m\to0$, one recovers General Relativity (GR) through the Vainshtein mechanism: The basic idea is to decouple the additional modes from the gravitational dynamics via nonlinear interactions of the helicity-0 mode of the graviton, \cite{Vainshtein:1972sx}. As a result, at the vicinity of matter, the non-linear interaction for the helicity-0 mode become large and hence suppresses its coupling to matter. This decoupling of the nonlinear helicity-0 mode is manifest in the limit where $M_4,M_5 \to \infty$ and $m\to 0$ while the strong coupling scale $\Lambda = (\mpl m^2)^{1/3}$ is kept fixed. This limit enables a linear treatment of the usual helicity-2 mode of gravity while the helicity-0 mode $\pi$ is described non-linearly, which is the so-called decoupling limit.

 One of the successes of the DGP model is the existence of a self-accelerating solution, where the acceleration of the Universe is sourced by the graviton own degrees of freedom (more precisely its helicity-0 mode).  Unfortunately that branch of solution seems to be plagued by ghost-like instabilities \cite{DDG,Koyama:2005tx,Charmousis:2006pn}, in the DGP model, but this issue could be avoided in more sophisticated setups, for instance including Gauss-Bonnet terms in the bulk \cite{deRham:2006pe}.

 More recently, it has been shown that the decoupling limit of DGP could be extended to more general Galilean invariant interactions \cite{Nicolis:2008in}. This Galileon model relies strongly on the symmetry of the helicity-0 mode $\pi$: Invariance under internal Galilean and shift transformations, which in induced gravity braneworld models can be regarded as residuals of the 5-dimensional Poincar\'e invariance. These symmetries and the postulate of ghost-absence restrict the construction of the effective $\pi$ Lagrangian. There exist only five derivative interactions which fulfill these conditions. From the five dimensional point of view these Galilean invariant interactions are consequences of Lovelock invariants in the bulk of generalized braneworld models, \cite{deRham:2010eu}. Since their inception there has been a flurry of investigations related to self-accelerating de Sitter solutions without ghosts \cite{Nicolis:2008in,Silva:2009km}, Galileon cosmology and its observations \cite{Chow:2009fm,Khoury:2009tk}, inflation \cite{Creminelli:2010ba,Burrage:2010cu,Mizuno:2010ag,Hinterbichler:2011qk}, lensing \cite{Wyman:2011mp}, superluminalities arising in spherically symmetric solutions around compact sources \cite{Hinterbichler:2009kq}, K-mouflage  \cite{kmouflage}, Kinetic Gravity Braiding \cite{Deffayet:2010qz}, etc... . Furthermore, there has been some effort in generalizing the Galileon to a non-flat background. The first attempt was then to covariantize directly the decoupling limit and to study its resulting cosmology \cite{Chow:2009fm}. In particular it was shown in \cite{Deffayet:2009wt} that the naive covariantization would yield ghost-like terms at the level of equation of motion but a given unique nonminimal coupling between $\pi$ and the curvature can remove these terms resulting in second order of equations of motion \cite{Deffayet:2009wt}, which are also consistent with a higher-dimensional construction \cite{deRham:2010eu}.
 In this paper, we will pursue the same strategy when constructing our proxy theory. The outcome of this kind of covariantization method was explored further in \cite{DeFelice:2010pv}.
 While this covariantization is ghost-free, the Galileon symmetry is broken explicitly in curved backgrounds. Only recently there has been a successful generalization to the (Anti-) de Sitter background and ultimately to maximally symmetric backgrounds where it has been discovered that the de Sitter Galileon interactions acquire additional, potential-like terms being functions of $\pi$ and mixtures of $\pi$ and gradients of $\pi$, all fulfilling a generalized Galileon symmetry \cite{Goon:2011qf}.

There exists a parallel to theories centered on a massive graviton: Galileon-type interaction terms naturally arise in gravitational theories using a massive spin-2 particle as an exchange particle, which has, in addition, been constructed to be ghost-free be it in three dimensions, \cite{deRham:2011ca} or for an generalized Fierz-Pauli action in four dimensions \cite{deRham:2010kj,deRham:2010ik}\footnote{Such a theory was also constructed using auxiliary extra dimensions, \cite{aux1,deRham:2010gu}. While in its most fundamental form, a ghost appears at quartic order in the decoupling limit, \cite{Hassan:2011zr}, it can also be cured order by order, \cite{Berezhiani:2011nc}.}.
Not only is the existence of a graviton mass a fundamental question from a theoretical perspective, it could also have important consequences both in cosmology and in solar system physics, \cite{Koyama:2011xz,Chkareuli:2011te}.
 Although solar system observations have confirmed GR to high accuracy and placed bounds on the graviton mass to be smaller than a few $\sim 10^{-32}$eV,  even such a small mass would become relevant at the Hubble scale which corresponds to the graviton Compton wavelength.
In particular, it has been successfully shown that this massive gravity theory exhibits a stable self-accelerated solution in the decoupling limit since the scalar mode can generate a constant negative pressure density. In the decoupling limit, the expansion history of the Universe in this self-accelerating branch was found to be indistinguishable from $\Lambda$CDM, \cite{deRham:2010tw}.

 While the self-accelerating solutions in the above models yield viable expansion histories including late-time acceleration, they do not address the cosmological constant problem, \ie the giant mismatch between the theoretically computed high energy density of the vacuum and the low observed value. A possible answer comes from the idea of degravitation, which asserts that the energy density could be as large as the theoretically expected value, but would not bear a large effect on the geometry. Technically, gravity is less strong on large scales (IR-limit) and could act as a high-pass filter suppressing the gravitational effect of a potentially large vacuum energy. Since such modifications of gravity in the IR naturally arise in models of massive gravity, they logically provide a possible mechanism to degravitate the vacuum energy density, \cite{degravitation,ArkaniHamed:2002fu,Dvali:2007kt}, which was observed in bi-Galileon models \cite{Padilla:2010tj} as well as in the decoupling limit of massive gravity, \cite{deRham:2010tw}. Analogously, the DGP braneworld model can be extended to higher dimensions to tackle the cosmological constant problem as well, \cite{Dvali:2007kt,gigashif,cascading}.

In this paper, we focus on the covariantization of a ghost-less extension of Fierz-Pauli massive gravity recently proposed in \cite{deRham:2010ik} and show that this proxy model allows for a stable self-accelerating solution. Hereby, we have performed the covariantization in the Jordan as well as in the Einstein frame. We discuss the differences between the two approaches and the consequences of choosing the wrong starting frame. In the well defined proxy theory we are able to tackle the puzzle about the self-acceleration of the Universe but not the one about the cosmological constant problem. Furthermore, we study the perturbations around the self-accelerating background and provide the expression for the modified evolution equation for these density perturbations. As expected, the $\pi$ field enhances the gravitational clustering resulting in a rapid growth of structures. This result is quite different from the one we had obtained in \cite{deRham:2010tw}, where the self-accelerating solution was indistinguishable from a $\Lambda$CDM, but we emphasize that the theory used in this paper, is a proxy model distinct from massive gravity. At early times, the enhancement of clustering is restrained since the Galileon self-interactions are the dominant ones suppressing their energy density relative to that of matter or radiation. Once the matter density has dropped sufficiently, the Galileon become an important contribution to the dark sector of the Universe driving cosmic expansion. \\

The paper is organized as follows: We start in section \ref{sec:MG1} with a summary of the non-linear theory of massive gravity \cite{deRham:2010ik} which we use for constructing our proxy theory, where we discuss the successful implementation of ghost-free massive gravity in the decoupling limit.
We then move to our proxy theory in section \ref{sec:proxy} where we first give some
motivations on why we use a proxy theory rather than the exact non-linear theory. We then work out the consequences of this covariantization for cosmology and specifically for late-time acceleration in section \ref{sec:SA} where we study also the stability conditions and the reason why a degravitating solution can not be constructed in our covariantized theory. We then move onto more general cosmology in section \ref{sec:cosmology} and follow the helicity-0 mode contribution to the Universe throughout its evolution, before quickly presenting the consequences for structure formation in section \ref{sec:Structure}.


\section{Massive Gravity and its decoupling limit}\label{sec:MG1}
We use the same notation as \cite{Nicolis:2008in}, with $(\p \pi)^2=g^{\mu\nu}\p_\mu \pi \p_\nu \pi$, $\Pi\mn=D_\mu D_\nu \pi$ and $\Pi^2_{\mu\nu}=g^{\alpha\beta}\Pi_{\mu\alpha}\Pi_{\beta\nu}$ where the covariant derivative is taken w.r.t. $g_{\mu\nu}$ and square brackets $[...]$ represent the trace of a tensor $[\Pi^2]=\Pi_{\mu\nu}\Pi^{\mu\nu}$ and $[\Pi]^2=\Pi^\mu_\mu\Pi^\nu_\nu$.

\subsection{Massive Gravity}
The first theory of massive gravity was proposed by Fierz and Pauli in 1939 \cite{FP}, but was shown to be unstable by Boulware-Deser (BD) \cite {BD}, due to the non-propagation of the Hamiltonian constraint at the non-linear level. This instability can also be seen in the \stu language \cite{AGS} where in the decoupling limit, the helicity-0 mode typically has equations of motion with more than two derivatives, and hence does not possess a well defined Cauchy surface.
 However it was shown in \cite{deRham:2010ik,deRham:2010kj} that the graviton potential could be built in such way as to remove any higher derivative term in the equations of motion, and obtain a Galileon-type of action for the additional helicity-0 mode. To review this, let us start with a graviton of mass $m$ described by
\ba
\mathcal L=\frac{\mpl^2}{2}\sqrt{-g}\left(R-\frac{m^2}{4}\mathcal U(g,H)\right)\label{eq:full}
\ea
with the potential $\mathcal U$, where the tensor $H\mn$ is constructed in terms of the metric $g\mn$ and the four \stu fields $\Phi^a$ by $H\mn=g\mn-\eta_{ab}\p_\mu \Phi^a \p_\nu \Phi^b$. We can then split the \stu fields into helicity-1 and -0 contributions, but the helicity-1 mode decouples in the decoupling limit and  can be consistently set to zero. We therefore focus on the helicity-0 mode $\pi$ and write  $\Phi^a=(x^a-\eta^{a\mu}\partial_\mu\pi)$ such that $H_{\mu\nu}=h_{\mu\nu}+2\Pi_{\mu\nu}-\eta^{\alpha\beta}\Pi_{\mu\alpha}\Pi_{\beta\nu}$.
Defining the quantity $\mathcal{K}^\mu_\nu(g,H)=\delta^\mu_\nu-\sqrt{\delta^\mu_\nu-H^\mu_\nu}$ the most generic potential that bears no ghosts in the decoupling limit is
\be
\mathcal U(g,H)=-4\(\mathcal{U}_2+\alpha_3\ \mathcal{U}_3+\alpha_4\ \mathcal{U}_4 \)\label{eq:fullU}
\ee
where $\alpha_{3,4}$ are two free parameters and
\ba
\mathcal{U}_2&=&[\mathcal{K}]^2-[\mathcal{K}^2]\\
\mathcal{U}_3&=&[\mathcal{K}]^3-3 [\mathcal{K}][\mathcal{K}^2]+2[\mathcal{K}^3]\\
\mathcal{U}_4&=&[\mathcal{K}]^4-6[\mathcal{K}^2][\mathcal{K}]^2+8[\mathcal{K}^3]
[\mathcal{K}]+3[\mathcal{K}^2]^2-6[\mathcal{K}^4]\,.
\ea
Notice that $\mathcal{U}_4$ can be expressed in terms of $\mathcal{U}_{2,3}$ and the tadpole $\mathcal{U}_1=[\mathcal{K}]$, \cite{Hassan:2011vm}.

It has then been shown that ghost-like pathologies in this theory of massive gravity theory disappear to all orders in the decoupling limit and at least up to quartic order beyond the decoupling limit, as well as completely non-linearly in some specific cases. While this theory is completely covariant, studying the cosmology as well as other non-trivial curved geometries can be extremely complicated. Instead, we will here focus on its version in the decoupling limit and covariantize the theory directly from this limit. We emphasize that the resulting proxy theory will be distinct from the theory of massive gravity presented above, but presents nevertheless some interesting features for cosmology.

In the decoupling limit, taking the scales $\mpl\to\infty$ and  $m\to0$, while keeping the strong coupling scale $\Lambda^3=\mpl m^2$ fixed, the above Lagrangian \eqref{eq:full} reduces to the more compact expression
\be
\mathcal L=-\frac12h^{\mu\nu}\mathcal E^{\alpha\beta}_{\mu\nu}h_{\alpha\beta}+h^{\mu\nu}X^{(1)}_{\mu\nu}+\frac{a_2}{\Lambda^3}h^{\mu\nu}X^{(2)}_{\mu\nu}+\frac{a_3}{\Lambda^6}h^{\mu\nu}X^{(3)}_{\mu\nu}+\frac{1}{2\mpl}h^{\mu\nu}T_{\mu\nu}\label{eq:massiveGravity}
\ee
where $h_{\mu\nu}$ stands for the helicity-2 mode canonically normalized, $\mathcal E^{\alpha\beta}_{\mu\nu}$ is the Lichnerowicz operator, the coefficients $a_{2,3}$ are related to the free parameters $\alpha_{3,4}$ and  $X^{(1,2,3)}\mn$ denote the interactions with the helicity-0 mode \cite{deRham:2010ik}
\ba
X^{(1)}_{\mu\nu} &=& \Box\pi g_{\mu\nu}-\Pi_{\mu\nu} \label{eq:X1} \\
X^{(2)}_{\mu\nu} &=& \Pi^2_{\mu\nu}-\Box\pi\Pi_{\mu\nu}-\frac12([\Pi^2]-[\Pi]^2)g_{\mu\nu}\\
X^{(3)}_{\mu\nu} &=& 6\Pi^3_{\mu\nu}-6[\Pi]\Pi^2_{\mu\nu}+3([\Pi]^2-[\Pi^2])\Pi_{\mu\nu}-g_{\mu\nu}([\Pi]^3-3[\Pi^2][\Pi]+2[\Pi^3])\,.\label{eq:X3}
\ea
It is worth to mention that these interaction terms are all transverse and at most second order in time derivatives to ensure the absence of ghost. Being ghost-free, these interactions are closely related to the Galileon interactions and fulfill the same internal symmetry. In the next section we will covariantize these interaction terms $h^{\mu\nu}X^{(1,2,3)}_{\mu\nu}$ and discuss their physical properties.\\
In the case of external sources there is a coupling between the metric $h_{\mu\nu}$ and the stress energy tensor via $h_{\mu\nu}T^{\mu\nu}$ but there is not such a direct coupling with the field $\pi$. Nevertheless, if one diagonalizes the first interaction term $h^{\mu\nu}X^{(1)}_{\mu\nu}$ by a change of variables of the form $h\mn=\bar h\mn+\pi \eta\mn$, then the coupling between $\pi$ and external sources will become transparent, $\pi T$.

\section{Massive Gravity: A proxy theory}\label{sec:proxy}
 Instead of studying the cosmology in the exact non-linear covariantized theory \eqref{eq:full} which can be extremely hard, we use the alternative approach of covariantizing the Lagrangian in the decoupling limit, and use the resulting theory as a proxy.
\subsection{Covariantization}
We claim that the covariantized version of the above Lagrangian \eqref{eq:massiveGravity}
 is simply given by
\be \label{eq:TotalAction}
S=\int\sqrt{-g}\left(\mpl^2 R+\mathcal{L}^\pi(\pi,g\mn)+\mathcal L^{\rm matter}(\psi,g\mn)\right)\,,
\ee
where $\mathcal{L}^{\rm matter}$ is the Lagrangian for the matter fields $\psi$ living on the geometry, the Lagrangian for $\pi$ is \cite{Chkareuli:2011te}
\be \label{eq:covJor}
\mathcal{L}^\pi=\mpl \(-\pi R-\frac{a_2}{\Lambda^3}\partial_\mu\pi\partial_\nu\pi G^{\mu\nu}-\frac{a_3}{\Lambda^6}\partial_\mu\pi\partial_\nu\pi \Pi_{\alpha\beta} L^{\mu\alpha\nu\beta}\)\,.
\ee
and the tensor $L^{\mu\alpha\nu\beta}$ stands for the dual Riemann tensor
\ba
L^{\mu\alpha\nu\beta}=2R^{\mu\alpha\nu\beta}+2(R^{\mu\beta}g^{\nu\alpha}+R^{\nu\alpha}g^{\mu\beta}-R^{\mu\nu}g^{\alpha\beta}-R^{\alpha\beta}g^{\mu\nu})
+R(g^{\mu\nu}g^{\alpha\beta}-g^{\mu\beta}g^{\nu\alpha})\,.
\ea
This form of tensor structure has been first discussed by Horndeski \cite{Horndeski} in the context of the most general scalar-tensor theory, and more recently in \cite{VanAcoleyen:2011mj,Charmousis:2011bf,Gubitosi:2011sg}. However we point out here that these interactions come as a direct outcome of massive gravity. We can show explicitly the following correspondences,
\ba
h^{\mu\nu}X^{(1)}_{\mu\nu} &\longleftrightarrow& -\pi R\\
h^{\mu\nu}X^{(2)}_{\mu\nu} &\longleftrightarrow& -\partial_\mu\pi\partial_\nu\pi G^{\mu\nu}\\
h^{\mu\nu}X^{(3)}_{\mu\nu}  &\longleftrightarrow& -\partial_\mu\pi\partial_\nu\pi \Pi_{\alpha\beta} L^{\mu\alpha\nu\beta}\,.
\ea\label{eq:corres}

The Einstein equation is given by
\be
\label{eq.:Einstein}
G_{\mu\nu}=\mpl T^\pi_{\mu\nu}+T^{\rm matter}_{\mu\nu}
\ee
with
\be
T^{\pi}_{\mu\nu}=T^{\pi(1)}_{\mu\nu}-\frac{a_2}{\Lambda^3}T^{\pi(2)}_{\mu\nu}-\frac{a_3}{\Lambda^6}T^{\pi(3)}_{\mu\nu}
\ee
and the structure of the Einstein and Riemann dual tensor ensure that $\pi$ enters at most with two derivatives in the stress-energy tensor,
\ba
T^{\pi(1)}_{\mu\nu}&=&X^{(1)}_{\mu\nu}+\pi G_{\mu\nu} \label{eq:T1} \nonumber\\
T^{\pi(2)}_{\mu\nu}&=&X^{(2)}_{\mu\nu}+\frac12L_{\mu\alpha\nu\beta}\partial^\alpha\pi\partial^\beta\pi+\frac12G_{\mu\nu}(\partial\pi)^2 \label{eq:T2} \nonumber\\
T^{\pi(3)}_{\mu\nu}&=&X^{(3)}_{\mu\nu}+\frac32L_{\mu\alpha\nu\beta}\Pi^{\alpha\beta}(\partial\pi)^2 \label{eq:T3}
\ea
where $X^{(i)}$ are defined in (\ref{eq:X1} - \ref{eq:X3}).
Furthermore, the fact that  $G^{00}$,  $G^{0i}$, $L^{0i0j}$ and $L^{0ikj}$  have at most one time-derivative guarantees the propagation of constraints.

Since we are not in the Einstein frame, these stress-energy tensor are only transverse on-shell, and satisfy the relation,  $D_\mu T^\mu_\nu=\partial_\nu\pi \mathcal E_\pi$ where $\mathcal E_\pi$ is the equation of motion with respect to $\pi$. Since both the Einstein tensor and the  Riemann dual tensor are transverse, this equation of motion is also at most second order in derivative,
\ba
\mathcal{E}_\pi&=&\frac{\delta \mathcal L^\pi}{\delta {\pi}}\nn\\
&=&-R-\frac{2a_2}{\Lambda^3}G^{\mu\nu}\Pi\mn-\frac{3a_3}{\Lambda^6}L^{\mu\alpha\nu\beta}(\Pi_{\mu\nu}\Pi_{\alpha\beta}+R^\gamma_{\;\;\beta\alpha\nu}\partial_\gamma\pi\partial_\mu\pi)=0\,,\label{eq:piJ}
\ea
where we have used the fact that
\ba
D_\nu D_\alpha D_\beta\pi L^{\mu\alpha\nu\beta}&=&-R^\gamma_{\;\;\beta\alpha\nu}\partial_\gamma\pi L^{\mu\alpha\nu\beta}=\frac14 \partial^\mu\pi \mathcal L_{\rm GB}
\ea
with $\mathcal L_{\rm GB}=R^2+R^2_{\mu\alpha\nu\beta}-4R^2_{\mu\nu}$.
In the rest of this paper, we study the resulting cosmology in this proxy theory, starting with the existence of self-accelerating solutions.
\section{de Sitter solutions}\label{sec:SA}
In what follows, we focus on the cosmology of the covariantized theory (\ref{eq:TotalAction}, \ref{eq:covJor}), and focus for that on a FRW background with scale factor $a(t)$ and Hubble parameter $H$.
The resulting effective energy density and pressure for the field $\pi$ are then
\ba
\rho^\pi&=&\mpl (6H\dot\pi+6H^2\pi-\frac{9a_2}{\Lambda^3}H^2\dot\pi^2-\frac{30a_3}{\Lambda^6}H^3\dot\pi^3)\label{eq:effdens}\\
P^\pi&=&3\mpl\left[\frac{6a_3}{\Lambda^6}H\dot\pi^2(\dot\pi(\dot H+H^2)+\frac32H\ddot\pi)+\frac{a_2}{2\Lambda^3}\dot\pi(\dot\pi(3H^2+2\dot H)+4H\ddot\pi)\right. \nonumber\\
&&\hspace{30pt} \left. -(\pi(3H^2+2\dot H)+2H\dot\pi+\ddot\pi)\right]\,,\label{eq:effpress}
\ea
and the equation of motion for $\pi$ (\ref{eq:piJ}) in the FRW space-time is equivalent to
\ba
\frac{6a_2}{\Lambda^3}\(3H^3\dot\pi+2H\dot H\dot\pi+H^2\ddot\pi\)+\frac{18a_3}{\Lambda^6}\(3H^2\dot H\dot\pi^2+3H^4\dot\pi^2+2H^3\dot\pi\ddot\pi\)=R.
\ea
Similarly as in \cite{Chow:2009fm}, this expression can be rewritten more compactly
\be
\ddot\phi+3H\dot\phi-R=0
\ee
if we define $\dot \phi$ as
\ba
\dot\phi=H^2\(\frac{6a_2}{\Lambda^3}\dot\pi+\frac{18a_3}{\Lambda^6}\dot\pi^2H\)\,.
\ea
\subsection{Self-accelerating solution}
Now, we would like to study the self-acceleration solution with $H=$const and $\dot H=0$. For the $\pi$ field we make the ansatz $\dot\pi=q\frac{\Lambda^3}{H}$. Furthermore, we assume that we are in a regime where $H \pi \ll \dot \pi$ so that we can neglect terms proportional to $\pi$ and consider only the terms including $\dot \pi$ or $\ddot \pi$. Thus, the Friedmann and field equations can be recast in
\ba
H^2=\frac{m^2}{3}(6q-9a_2 q^2-30a_3 q^3)\label{eq:modFrid}\\
H^2(18a_2 q+54a_3 q^2-12)=0\,.
\ea
Assuming $H\not=0$, the field equation then imposes,
\be
q=\frac{-a_2\pm\sqrt{a_2^2+8a_3}}{6a_3}
\ee
while the Friedmann equation (\ref{eq:modFrid}) sets the Hubble constant of the self-accelerated solution. Similar to what has been found in \cite{deRham:2010tw} our proxy theory admits a self-accelerated solution, with the Hubble parameter set by the graviton mass.
For the stability condition of this self-accelerating solution the first constraint we have is to demand $H>0$. The other constrain comes from the stability condition for perturbations on the background which we discuss in the following subsection.

\subsection{Stability conditions}
In the last subsection we have shown explicitly that our proxy theory exhibits a self-accelerating solution with $H^2\approx m^2$. Now, we would like to study whether the perturbations on this background are stable and what the constraints are. For this purpose, consider perturbations on the background solution of the following form
\be
\pi=\pi_0(t)+\delta\pi(t,x,y,z)\label{eq:perturbations}
\ee
The second order action for the perturbations is
\be
\mathcal L=-\frac{a_2}{\Lambda^3}\partial_\mu\delta\pi\partial_\nu\delta\pi G^{\mu\nu}-\frac{a_3}{\Lambda^6}\partial_\mu\delta\pi\partial_\nu\delta\pi\Pi^{(0)}_{\alpha\beta}L^{\mu\alpha\nu\beta}-\frac{2a_3}{\Lambda^6}\partial_\mu\pi_0\partial_\nu\delta\pi D_\alpha D_\beta\delta\pi L^{\mu\alpha\nu\beta}\label{eq:perLag}\,,
\ee
which can be written in the form
\be
\mathcal L=K_{tt}(\delta\dot\pi^2-\frac{c_s^2}{a^2}(\nabla \delta\pi)^2)
\ee
where
\be
K_{tt}=-\frac{3\mpl a^3H^2}{\Lambda^3}\left(a_2+\frac{6a_3 H}{\Lambda^3}\dot\pi\right)
\ee
and
\be
c_s^2=\frac13 \left(2+\frac{a_2\Lambda^3}{a_2\Lambda^3+6 a_3 H\dot\pi}\right)\,.
\ee
The condition for the stability is then given by $K_{tt}>0$, $c_s^2>0$ and $H^2>0$, which are fulfilled if
\be\label{eq:stability}
a_2>0 \;\;\;\;\;\;\;{\rm and}\;\;\;\;\;\; 0>a_3>-\frac18 a_2^2.
\ee
To compare this result with the condition obtained in the decoupling limit, \cite{deRham:2010tw}, we first mention that $a_2^{{\rm here}}=-2a_2^{{\rm there}}$ and $a_3^{{\rm here}}=a_3^{{\rm there}}$. In terms of the parameters used there, we need to compare our conditions $a_2^{{\rm here}}<0$, $a_3^{{\rm here}}>-\frac18 (a_2^{{\rm here}})^2$ to the conditions $a_2^{{\rm there}}<0$ and $-\frac23(a_2^{{\rm there}})^2<a_3^{{\rm there}}<-\frac12(a_2^{{\rm there}})^2$.  We see that our theory is less constraining but is still within the parameter space derived in \cite{deRham:2010tw}. It is not surprising that the stability condition in the decoupling limit and in our covariantized theory do not coincide totally as we have explicitly broken the symmetry when getting the proxy and in particular, our solution does now spontaneously break Lorentz invariance, which was not the case in the decoupling limit, \cite{deRham:2010tw}.

It is also worth pointing out that the self-accelerating solution by itself does not propagate any superluminal mode, since $2/3<c_s^2<1$.

We emphasize as well that the constant $\dot\pi$ solution is a dynamical attractor. For this we just consider time dependent perturbations $\pi(t)=\pi_0(t)+\delta \pi(t)$ which is a special case of \eqref{eq:perturbations} fulfilling the same stability conditions.
The equation of motion for perturbations simplifies to
\be
\partial_t(a^3 \delta \dot \pi)=0
\ee
The solution for $\delta \dot\pi$ is given by
\be
\delta \dot\pi(t)\sim a^{-3}\,.
\ee
Thus, these perturbations $\delta \pi(t)$ redshift away exponentially compared to the $\dot\pi={\rm const}$ self-accelerating solution. Therefore, the self-accelerating solution is an attractor.

\subsection{Degravitation}
More interestingly, one can wonder whether degravitation can be exhibited in these class of solutions. If one take $\pi=\pi(t)$ and $H=0$, it is straightforward to see that we  obtain $\rho^\pi=0$, so the field has absolutely no effect and cannot help the background  to degravitate. The situation is however different when setting $\pi=\chi_0 x_\alpha x^\alpha$, which was possible in the decoupling limit. In this proxy theory,
such a behaviour will involve explicit space dependances in the equations of motion which should be dealt with specifically.

Interestingly, the interactions considered here are precisely of the same form as that studied recently in \cite{Charmousis:2011bf}. There as well, in the absence of spatial curvature $\kappa=0$, the contribution from the scalar field vanishes if $H=0$. Comparing with \cite{Charmousis:2011bf}, we can hence wonder whether the addition of spatial curvature $\kappa\ne 0$ in our proxy theory could help achieving degravitation, but relying strongly on spatial curvature brings concerns over instabilities which are beyond the scope of this study.

\section{Cosmology}\label{sec:cosmology}

In the following we would like to discuss in more detail the interplay of all the constituents of the universe. We assume that matter, radiation and the scalar field $\pi$ contribute to the total energy density of the universe.
\be
H^2=\frac{8\pi G}{3}(\rho^\pi +\rho^{rad} +\rho^{mat})
\ee
Consider the scalar field $\pi$ as a perfect fluid with the effective energy density and pressure given by (\ref{eq:effdens}, \ref{eq:effpress}). Thus, the equation of state parameter of this new field would be
\ba
\omega_\pi&=&\frac{-12a_3H^3\dot\pi^3+2\Lambda^3\dot H(2\Lambda^3\pi-a_2\dot\pi^2)+2\Lambda^6\ddot\pi+4H\dot\pi(\Lambda^6-3a_3\dot H\dot\pi^2-a_2\Lambda^3\ddot\pi)}{3H(2\Lambda^6\dot\pi-10a_3H^2\dot\pi^3+H(2\Lambda^6\pi-3a_2\Lambda^3\dot\pi^2))}\nonumber\\
&&+\frac{3H^2(2\Lambda^6\pi-\dot\pi^2(a_2\Lambda^3+6a_3\ddot\pi))}{3H(2\Lambda^6\dot\pi-10a_3H^2\dot\pi^3+H(2\Lambda^6\pi-3a_2\Lambda^3\dot\pi^2))}\,.
\ea
At this point one should mention that the energy density for the $\pi$-field is not conserved but rather given by $D_\mu T^\mu_\nu=\partial_\nu\pi \mathcal E_\pi$ (where $\mathcal E_\pi$ is the equation of motion for $\pi$), which is not surprising since $\pi$ is non-minimally coupled to gravity in the Jordan frame. Therefore, we can have $\omega_\pi\textless-1$.\\
In the following  we will first assume that at early times in the evolution history of the Universe we can neglect the extra density coming from the helicity-0 $\rho_\pi$. We will then check this assumption by plugging the solution for $H$ back in the equation of motion for $\pi$.
If we assume that at early times the radiation density dominates, we simply have
\be
H^2=\frac{8\pi G}{3}\rho_0^{\rm rad}a^{-4} \;\;\;\;\;\;\;a\sim t^{1/2}\;\;\;\;\;\;\;\;\;\;\;\omega=1/3\label{eq:raddom}
\ee
 During the radiation era, the dominant terms in the equation of motion for $\pi$ are then  $\frac{54a_3}{\Lambda^6}H^2\dot H\dot\pi^2+\frac{54a_3}{\Lambda^6}H^4\dot\pi^2+\frac{36a_3}{\Lambda^6}H^3\dot\pi\ddot\pi=0$ which can be solved assuming the previous expression for $H$ (\ref{eq:raddom})
\be
\pi_{\rm rad}\sim t^{1.75}\;\;\;\;\;\; {\rm yielding} \;\;\;\;\;\; \rho^{\pi}_{\rm rad}\sim \mpl t^{-1/4}
\ee
\begin{figure}[h]
\begin{center}
 \includegraphics[width=0.80\textwidth]{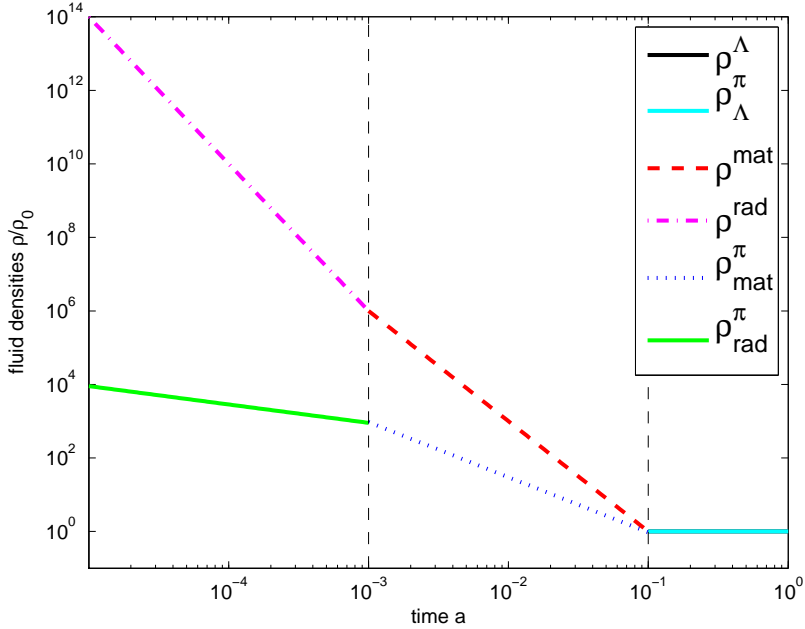}
  \caption{Fluid densities  $\rho^{rad}\sim a^{-4}$, $\rho^{mat}\sim a^{-3}$ and $\rho_\pi$ during the epochs of radiation, matter and $\Lambda$-domination normalised to today $\rho_\pi$. During the radiation domination the energy density for $\pi$ goes as $\rho^\pi_{\rm rad}\sim a^{-1/2}$ and during matter dominations as $\rho^\pi_{\rm mat}\sim a^{-3/2}$ and is constant for later times $\rho^\pi_\Lambda={\rm const}$.}
\end{center}
\end{figure}
\\
At later times when the matter dominated epoch starts we have
\be
H^2=\frac{8\pi G}{3}\rho_0^{\rm mat}a^{-3} \;\;\;\;\;\;\;a\sim t^{2/3}\;\;\;\;\;\;\;\;\;\;\;\omega=0
\ee
Now the dominant terms in the equation of motion for $\pi$ are $\frac{18a_2}{\Lambda^3}H^3\dot\pi+\frac{12a_2}{\Lambda^3}H\dot H\dot\pi+\frac{6a_2}{\Lambda^3}H^2\ddot\pi-12H^2-6\dot H=0$. We get for $\pi$ this time
\ba
\pi^{\rm mat}\sim c_2\cdot t+\frac{t^2\Lambda^3}{4a_2}\;\;\;\;\;\;{\rm yielding} \;\;\;\;\;\; \rho^\pi_{\rm mat}=c_2 \mpl t^{-1}+ \frac{3\mpl(-14a_2^2+5a_3)\Lambda^3}{32 a_2^3}\,.
\ea
Summarizing, during radiation domination the effective energy density for the $\pi$-field goes like $\rho^\pi_{\rm rad} \sim t^{-1/4}$ while during matter domination as $\rho^\pi_{\rm mat} \sim t^{-1}$ and approaches a constant at late time. As shown in the figure $\rho^\pi$ can be neglected at early times where $\rho^\pi_{\rm mat}\ll\rho^{\rm mat}$ and $\rho^\pi_{\rm rad}\ll\rho^{\rm rad}$.

\section{Structure formation}\label{sec:Structure}
We end the cosmological analysis by looking at the evolution of matter density perturbations. The density perturbations follow the evolution
\be
\ddot \delta_m+2H\dot\delta_m=\frac{\nabla^2\psi}{a^2}
\ee
where $\psi$ is the Newtonian potential. The effects of $\pi$ are all encoded in its contribution to the Poisson equation. Consider perturbations of $\pi$ around its cosmological background solution $\pi(x,t)=\pi_0(t)+\phi(x,t)$. $\phi$ gives a contribution to the Newtonian potential of the form $\psi=\phi/\mpl$.
In the Newtonian approximation we have $|\dot\phi|\ll|\nabla\phi|$. The equation of motion for the scalar field in first order in $\phi$ is
\be
-\frac{2a_2}{\Lambda^3}G^{\mu\nu} D_\mu D_\nu\phi-\frac{2a_3}{\Lambda^3} L_{\mu\alpha\nu\beta}(4\Pi_{0}^{\alpha\beta} D^\mu D^\nu\phi+2R^{\gamma\beta\alpha\nu}\partial_\gamma\pi_{0}\partial^\mu\phi)=\delta R
\ee
which is equivalent to (neglecting $\dot\phi$)
\ba
\left[-\frac{2a_2}{\Lambda^3}(3H^2+2\dot H)+\frac{16a_3}{\Lambda^6}(2H^3 \dot \pi_0+2 \dot H H \dot \pi_0+H \ddot \pi_0)\right]\frac{\nabla^2\phi}{a^2}=\delta R
\ea
and last but not least we need the trace of Einstein equation, (\ref{eq.:Einstein}-\ref{eq:T3}). Perturbing the trace to first order, we get
\ba
-\mpl\delta R&=&\Big[3-2 \frac{a_2}{\Lambda^3}(2H \dot \pi_0+\ddot \pi_0)-\frac{3a_3}{\Lambda^6}(2\dot H \dot \pi^2_0+5 H^2 \dot \pi_0^2+4 H \ddot \pi_0 \dot \pi_0)
\Big]\frac{\nabla^2\phi}{a^2}\nn \\
&&+
\frac{\delta T}{\mpl}
\ea
 To reach that point, we have neglected the perturbations of the curvature of the form $\delta R \pi_0$ as they are negligible compared to $\mpl \delta R$ since we work in the regime where $\pi_0\ll\mpl$. We have also ignored terms of the form $\delta R (\partial \pi_0)^2/\Lambda^3$, which could a priori be relevant, but their inclusion would require solving the full Einstein equation, which is beyond the scope of this study. As a first approximation, such terms are hence ignored.

The perturbations for the source is just given by $\delta T=-\rho_m\delta_m$ for non-relativistic sources, thus we have
\be
\frac{\nabla^2\phi}{a^2}=\frac{\rho_m\delta_m}{3 M_{\rm Pl}\mathcal Q}
\ee
where $\mathcal Q$ stands for
\ba
&&\hspace{-10pt}\mathcal Q\equiv 1-\frac{2 a_2}{\Lambda^3}(2 H \dot \pi_0+\ddot \pi_0+\mpl (2 \dot H + 3 H^2))\\
&&-\frac{a_3}{\Lambda^6}\(5H^2 \dot \pi^2_0+2 \dot H \dot pi_0^2+4 H \ddot \pi_0 \dot \pi_0-\frac{16 \mpl}{3}(2 H^3 \dot \pi_0+2 H \dot H \dot \pi_0+H^2 \ddot \pi_0)\)\nn\,.
\ea
Finally, the modified evolution equation for density perturbations is
\be
\ddot \delta_m+2H\dot\delta_m=\frac{\rho_m\delta_m}{M^2_{\rm Pl}}\left(1+\frac1{3\mathcal Q}\right)\,.
\ee
Knowing the background configuration it is then relatively straightforward to derive the effect on structure formation. We recover the usual result that when the field is screened, $H \dot \pi \gtrsim \Lambda^3$, the extra force coming from the helicity-0 is negligible and the formation  of  structure is similar as in $\Lambda$CDM.

\section{Covariantization from the Einstein Frame}\label{sec:Einstein}
After having studied the cosmology and the structure formation of our proxy theory in the Jordan frame, the natural question is whether we would expect similar results in a different  frame. Instead of covariantizing our Lagrangian in the  Jordan frame, it is on an equal footing to go to the Einstein frame where the Ricci scalar is not multiplied by the scalar field $\pi$ and covariantize the theory at that stage. Since it is unclear at first sight which frame is the physical frame we will consider both frames and discuss their differences. Our starting Lagrangian was
\be
\mathcal L=-\frac12h^{\mu\nu}\mathcal E^{\alpha\beta}_{\mu\nu}h_{\alpha\beta}+h^{\mu\nu}X^{(1)}_{\mu\nu}+\frac{a_2}{\Lambda^3}h^{\mu\nu}X^{(2)}_{\mu\nu}+\frac{a_3}{\Lambda^6}h^{\mu\nu}X^{(3)}_{\mu\nu}+\frac{1}{2\mpl}h^{\mu\nu}T_{\mu\nu}
\ee
Now, when we do the following change of variables
\be
h_{\mu\nu}=\bar h_{\mu\nu}+\pi\eta_{\mu\nu}\label{eq:EF1}
\ee
we can diagonalize the first mixed term $h^{\mu\nu}X^{(1)}_{\mu\nu}$ such that the Lagrangian takes the following form
\ba
\mathcal L=-\frac12\bar h^{\mu\nu}\mathcal E^{\alpha\beta}_{\mu\nu}\bar h_{\alpha\beta}+\frac32\pi\Box\pi+\frac{a_2}{\Lambda^3}\bar h^{\mu\nu}X^{(2)}_{\mu\nu}-\frac32\frac{a_2}{\Lambda^3}\Box\pi(\partial\pi)^2\nonumber\\
 +\frac{a_3}{\Lambda^6}\bar h^{\mu\nu}X^{(3)}_{\mu\nu}
-2\frac{a_3}{\Lambda^6}(\partial\pi)^2([\Pi^2]-\Box\pi^2)  +\frac{1}{2\mpl}(\bar h_{\mu\nu}+\pi\eta_{\mu\nu})T^{\mu\nu}
\ea
Covariantizing this action is straightforward. We use again the correspondences in \eqref{eq:corres} and it has been shown explicitly that the covariant equivalence to $\Box\pi(\partial\pi)^2$ and $-2(\partial\pi)^2([\Pi^2]-\Box\pi^2)$ are given by $\Box\pi(\partial\pi)^2$ and $2(\partial\pi)^2(\Box\pi^2-[\Pi^2]-\frac14(\partial\pi)^2R) $ respectively, which do not yield any ghostlike instabilities  (\cite{Deffayet:2009wt}, \cite{deRham:2010eu}). Thus, the covariantized action in the Einstein frame is simply given by
\ba
\mathcal{L}&=&\mpl^2 R+\frac32\pi\Box\pi-\frac{a_2\mpl}{\Lambda^3}\partial_\mu\pi\partial_\nu\pi G^{\mu\nu}-\frac32\frac{a_2}{\Lambda^3}\Box\pi(\partial\pi)^2\nonumber\\
&&-\frac{a_3\mpl}{\Lambda^6}\partial_\mu\pi\partial_\nu\pi \Pi_{\alpha\beta} L^{\mu\alpha\nu\beta}
+2\frac{a_3}{\Lambda^6}(\partial\pi)^2(\Box\pi^2-[\Pi^2]-\frac14(\partial\pi)^2R)  \label{eq:einstein}\\
&&+\mathcal L_m(\psi, (1+\pi)g_{\mu\nu})\nn
\ea
Similarly as before, the properties of $G^{\mu\nu}$ and $L^{\mu\alpha\nu\beta}$ ensure that their equations of motion lead at most to second order derivative terms.
To find a self-accelerating solution we set again a pure de Sitter metric, with $\dot\pi=q\Lambda^3/H$.  The Friedmann and the field equations then take the form
\ba
&&\mpl^2H^2=3\mpl a_2\Lambda^3q^2+10\mpl a_3\Lambda^3q^3-\frac12\frac{\Lambda^6}{H^2}q^2+3a_2\frac{\Lambda^6}{H^2}q^3+15a_3\frac{\Lambda^6}{H^2}q^4\nonumber\\
&& \Lambda^3(-1+3q(a_2+4a_3q))+2\mpl(a_2+3a_3q)H^2=0\label{eq:FRein}
\ea
When comparing the above Friedmann and the field equations with the one we had in the Jordan frame \eqref{eq:modFrid}, we see significant differences coming from the extra terms which were not there in the Jordan frame. These terms yield Friedmann and field equations proportional to $q^4$ and $H^4$ which are more difficult to solve.
\\
 For fairness, we should compare both actions in the same frame. We do so by performing a conformal transformation on the action \eqref{eq:covJor}:
\be
\tilde g_{\mu\nu}=\Omega^2 g_{\mu\nu}\;\;\;\;\;\; {\rm with} \;\;\;\;\;\;\; \Omega^2=\left(1-\frac{\pi}{\mpl}\right)\,.
\ee
For simplicity we consider the case for which $a_3=0$, so under this conformal transformation the covariantized action \eqref{eq:covJor} becomes
\ba
\mathcal{L}_J= \mpl^2\tilde R-\frac32\Omega^{-4}(\tilde \partial \pi)^2-\frac{a_2\mpl}{\Lambda^3}\left(\tilde\partial^\mu\pi\tilde\partial^\nu\pi \tilde G_{\mu\nu}+\frac32\frac{\Omega^{-2}}{\mpl}(\tilde\partial\pi)^2\tilde\Box\pi+\frac54\frac{\Omega^{-4}}{\mpl^2}(\tilde \partial \pi)^4\right)
\ea
In the limit where $\pi\ll\mpl$ we have then finally the following expression
\ba
\mathcal{L}_J= \mpl^2 R+\frac32\pi\Box\pi-\frac{a_2\mpl}{\Lambda^3}\partial_\mu\pi\partial_\nu\pi G^{\mu\nu}-\frac32\frac{a_2}{\Lambda^3}\Box\pi(\partial\pi)^2\,,
\ea
which coincides with the theory obtained from the Einstein frame, \eqref{eq:einstein} with $a_3=0$. So within the regime of validity of our results, our conclusions are independent of the choice of frame. However beyond this regime the theory originally constructed from the Jordan frame could violate the null energy condition from the term proportional to $(\partial\pi)^4$.

\section{Discussion}\label{sec:conclusion}

Common to generic massive gravity models is the presence of at least one additional helicity-0 degree of freedom which originates from a spin-two graviton field. In contrast to an arbitrary scalar field, its dynamics is not driven by a potential but rather by specific derivative terms fixed by symmetries. Such a degree of freedom is experimentally testable. On solar system and galactic scales gravity is very well compatible with GR and correctional terms from a modified gravity model are excluded. On these scales, however, the effects of massive gravity might be cloaked by the Vainshtein mechanism, where the helicity-0 mode interactions become appreciable to freeze out the field fluctuations. A number of screening mechanisms have been devised such as the chameleon and symmetron mechanisms. Contrary to those the success of the Vainshtein mechanism does not rely on some specific potential but instead on derivative interactions, which cause the helicity-0 mode to become decoupled from matter and light on short distances, which nevertheless could have observational signatures on larger scales in cosmic structure formation.

A further benefit of the Vainshtein mechanism would be a natural solution to the problem that the matter density and the cosmological constant are of similar magnitudes today. The cosmological application of the Vainshtein mechanism works such that at early times, the galileon interactions are dominated by self-interaction, which suppresses their energy density relative to that of matter or radiation. If the matter density has decreased sufficiently by cosmic expansion, the galileon constitutes an important contribution to the energy density of the universe and drives cosmic expansion.

Although the precise cosmology of the theory of massive gravity proposed in \cite{deRham:2010kj} has not been derived,  most  models of massive gravity give rise to alterations of the Hubble function similar to DGP-gravity. Experimental tests of the expansion dynamics of the Universe include the distance-redshift relation of supernovae, and measurements of the angular diameter distance as a function of redshift, as in the case of the cosmic microwave background and the baryon acoustic oscillations. Apart from these geometrical tests, the time-dependence of evolving cosmic structures can be investigated, and the influence of the gravitational theory on the geodesics of relativistic (photons) and nonrelativistic (dark matter) test particles. The first category includes gravitational lensing and the Sachs-Wolfe effects, which have been shown to differ from their GR-expectation in some modified gravity theories, and a similar result can be expected in massive gravity. The second category includes the homogeneous growth of the cosmic structure, and the formation of galaxies and clusters of galaxies by gravitational collapse. Again, massive gravity would influence the time sequence of gravitational clustering and the evolution of peculiar velocities, as well as the number density of collapsed objects. In particular, massive gravity would enhance gravitational clustering since they tend to lower the collapse threshold for density fluctuations in the large-scale structure, leading to a higher comoving number density of galaxies and clusters of galaxies. Naturally, these changes are degenerate with a different choice of cosmological parameters and with introducing non-Gaussian initial conditions, which would be very interesting to quantify. Recent discrepancies of $\Lambda$CDM with observational data on large scales include the number of very massive clusters, the strong lensing cross section, anomalous multipole moments of the CMB, the axis of evil, and large peculiar velocities. It is beyond the scope of this paper to address these issues using massive gravity, but we propose how to proceed from constructing a massive gravity theory to providing observationally testable quantities. From our point of view it is advisable to focus on probes of large scales, due to difficulties related to nonlinear structure formation and the influence of baryons on small scales. Natural questions concern the homogeneous dynamics of the Universe, the formation of structures and the shape of geodesics of relativistic and nonrelativistic particles. Basically, using our proxy theory one should be able to make predictions concerning these four issues and the combination of the four should give insight into the nature of the gravitational sector. In our proxy theory, from the modified field equation the Hubble-function can be derived easily, which allows the definition of distance measures, needed in the interpretation of supernova data. Cosmic structure growth tests the Newtonian limit for slowly-moving particles and describes the clustering of galaxies and the growth of structures which are investigated by e.g. gravitational light deflection. Lensing, in turn, makes use of the geodesic equation for relativistic particles, and measures the correlation function of the matter density, weighted with the lensing efficiency function, which in turn is derived from distance measures. It would be quite interesting to study these observational consequences and constraints of our proxy theory in a future work.

\section*{Acknowledgments}

We would like to thank Clare Burrage, Paul de Fromont, Gregory Gabadadze, Parvin Moyassari and Andrew Tolley for useful discussion. Furthermore, special thanks for Robert Caldwell!
This work is supported by the SNF.

\end{document}